\documentclass[12pt]{article}

\usepackage{epsfig}
\usepackage{amssymb}
\usepackage{amsmath}
\usepackage{color}
\usepackage{enumitem, array}
\usepackage{hyperref}
\usepackage{bbm}
\usepackage{mathrsfs}
\usepackage{colonequals}
\usepackage{cite}
\newcommand{\beqa}{\begin{eqnarray}}
\newcommand{\eeqa}{\end{eqnarray}}
\newcommand{\beq}{\begin{equation}}
\newcommand{\eeq}{\end{equation}}

\makeatletter \@addtoreset{equation}{section} \makeatother

\newdimen\yysquaresize
\newdimen\yyrsquaresize
\newdimen\yythickness
\newdimen\yyskip
\def\yysquare#1{%
\setlength{\yyrsquaresize}{\yysquaresize}%
\addtolength{\yyrsquaresize}{-2\yythickness}%
\vrule width \yythickness%
\vbox to \yysquaresize{%
  \hrule height \yythickness\vss%
  \hbox to \yyrsquaresize{\hss#1\hss}%
  \vss\hrule height \yythickness}%
\vrule width \yythickness}

\newcounter{multieqs}

\newcommand{\be}{\begin{equation}}
\newcommand{\ee}{\end{equation}}
\newcommand{\bea}{\begin{eqnarray}}
\newcommand{\eea}{\end{eqnarray}}

\def\g{{\mathfrak{g}}}

\def\be{\begin{equation}}
\def\ee{\end{equation}}
\def\bea{\begin{eqnarray}}
\def\eea{\end{eqnarray}}

\def\nn{\nonumber}

\def\ph{\phantom}

\global \long \def \l{\lambda}

\global \long \def \r{\rho}

\global \long \def \Nm{\mathcal{N}}

\global \long \def \Bm{\mathcal{B}}
\global \long \def \Cm{\mathcal{C}}

\global \long \def \Pm{\mathcal{P}}

\global \long \def \Lm{\mathcal{L}}

\global \long \def \Am{\mathcal{A}}

\global \long \def \a{\alpha}
\global \long \def \b{\beta}
\global \long \def \g{\gamma}

\global \long \def \ad{\dot{\alpha}}
\global \long \def \bd{\dot{\beta}}
\global \long \def \gd{\dot{\gamma}}

\global \long \def \ab{\textbf{a}}
\global \long \def \bb{\textbf{b}}

\global \long \def \dag{\dagger}

\global \long \def \ph{\phantom}

\global \long \def \dag{\dagger}
\global \long \def \pd{\partial}

\def\Nm{{\mathcal{N}}}
\def\Om{{\mathcal{O}}}
\def\Bm{{\mathcal{B}}}
\def\Cm{{\mathcal{C}}}

\def\veps{\varepsilon}
\def\jb{\bar{\jmath}}
\def\ph{\phantom}
\def\pd{\partial}

\def\nn{\nonumber}

\def\be{\begin{equation}}
\def\ee{\end{equation}}
\def\bea{\begin{eqnarray}}
\def\eea{\end{eqnarray}}

\begin{document}

\thispagestyle{empty}
\setcounter{page}{0}
\begin{flushright}\footnotesize
\texttt{DESY 17-002}
\vspace{0.5cm}
\end{flushright}
\setcounter{footnote}{0}

\begin{center}%
{\Large\textbf{\mathversion{bold}%
Revisiting the dilatation operator of the\\ Wilson-Fisher fixed point }\par}

\vspace{15mm}
{\sc 
 Pedro Liendo}\\[5mm]

{\it DESY Hamburg, Theory Group, Notkestrasse 85, D–22607 Hamburg, Germany
}\\[5mm]

\small{\texttt{pedro.liendo@desy.de}}\\[20mm]

\textbf{Abstract}\\[2mm]
\end{center}
We revisit the order $\varepsilon$ dilatation operator of the Wilson-Fisher fixed point obtained by Kehrein, Pismak, and Wegner in light of recent results in conformal field theory. Our approach is algebraic and based only on symmetry principles. The starting point of our analysis is that the first correction to the dilatation operator is a conformal  invariant, which implies that its form is fixed up to an infinite set of coefficients associated with the scaling dimensions of higher-spin currents. These coefficients can be fixed using well-known perturbative results, however, they were recently re-obtained using CFT arguments without relying on perturbation theory. Our analysis then implies that all order-$\varepsilon$ scaling dimensions of the Wilson-Fisher fixed point can be fixed by symmetry.

\newpage

\setcounter{tocdepth}{2}
\hrule height 0.75pt
\tableofcontents
\vspace{0.8cm}
\hrule height 0.75pt
\vspace{1cm}

\setcounter{tocdepth}{2}

\section{Introduction}
\label{sec:introduction}

The Wilson-Fisher fixed point is one of the paradigmatic examples of conformal field theory (CFT). It can be obtained from a free scalar in $d=4-\veps$ dimensions by adding a quartic interaction and flowing to the IR. The usual approach is to consider a perturbative expansion in the parameter $\veps$, with the hope that setting $\veps=1$ will give an approximate description of three-dimensional physics \cite{Wilson:1971dc}. The continuation of the Wilson-Fisher fixed point to $d=3$ belongs to the same universality class of the critical Ising model, and indeed, critical exponents obtained using the $\veps$-expansion are in good agreement with results obtained by other methods \cite{Pelissetto:2000ek}.

The main motivation behind this work is the bootstrap approach to CFTs \cite{Ferrara:1973yt,Polyakov:1974gs,Belavin:1984vu}, which attempts to constrain the dynamics of a theory using only the conformal algebra. This approach has experienced a revival thanks to the pioneering numerical work of \cite{Rattazzi:2008pe}, that has motivated many recent papers on the bootstrap, but also influenced research on related areas, like mathematical physics, supersymmetric field theory, and holography. Of all these recent developments, perhaps the most impressive is the high-precision estimates of critical exponents in the three-dimensional Ising model \cite{El-Showk:2014dwa,El-Showk:2013nia,Kos:2014bka,Simmons-Duffin:2015qma,Kos:2016ysd}.

The new developments have also led to revisit the $\veps$-expansion through the lens of the bootstrap.\footnote{For interesting related work that considers a large-charge expansion of CFTs see \cite{Alvarez-Gaume:2016vff,Hellerman:2015nra}.} This approach was first considered in the context of boundary CFT, where an attempt was made to recover the $\veps$-expansion starting from crossing symmetry \cite{Liendo:2012hy}. A different strategy not based in crossing was presented in \cite{Rychkov:2015naa} (see also \cite{Basu:2015gpa,Ghosh:2015opa,Raju:2015fza,Nii:2016lpa,Bashmakov:2016pcg,Yamaguchi:2016pbj,Hasegawa:2016piv,Roumpedakis:2016qcg}), where the constraints imposed by multiplet recombination turned out to be powerful enough to fix the order $\veps$ anomalous dimensions of operators of the form $\phi^n$. Three more techniques include  using the singularity structure of conformal blocks \cite{Gliozzi:2016ysv,Gliozzi:2017hni}, unitarity methods \cite{Sen:2015doa}, and the Mellin bootstrap \cite{Gopakumar:2016wkt,Gopakumar:2016cpb,Dey:2016mcs}, where corrections up to order $\veps^3$ have been obtained.

Another approach based on the equations of motion and similar in spirit to \cite{Rychkov:2015naa} was used in \cite{Giombi:2016hkj,Skvortsov:2015pea} (see also \cite{Giombi:2017rhm}), where the leading order anomalous dimensions of higher-spin currents were obtained. In particular, the order-$\veps$ anomalous dimensions vanish, which will play a crucial role in our analysis. The same result was obtained without recourse to a Lagrangian description in \cite{Alday:2016njk,Alday:2016jfr} using the large-spin limit of crossing symmetry \cite{Fitzpatrick:2012yx,Komargodski:2012ek}, and in \cite{Roumpedakis:2016qcg} using a generalization of the method of \cite{Rychkov:2015naa}.

The purpose of this paper is to study the complete order $\veps$ dilatation operator of the Wilson-Fisher fixed point, which allows to efficiently calculate the anomalous dimension of any operator in the theory. This result is not new, it was obtained by a careful analysis of Feynman diagrams by  Kehrein, Pismak, and Wegner in \cite{Kehrein:1992fn}. They packaged their result in an elegant formula and checked a posteriori that their expression was invariant under the conformal algebra. Here we will reverse the logic, and consider invariance of the dilatation operator as our starting point, as shown below, this drastically simplifies the calculation. This approach was successfully used by Beisert to obtain the complete dilatation operator of $\Nm=4$ SYM \cite{Beisert:2003jj,Beisert:2004ry}. The technique however is quite general and not restricted to superconformal theories. It turns out that the few coefficients that cannot be fixed by Beisert's approach, have been recently obtained using the aforementioned symmetry arguments \cite{Alday:2016njk,Alday:2016jfr,Roumpedakis:2016qcg}. Thus, the dilatation operator of the Wilson-Fisher fixed point can be completely fixed by symmetry, bypassing Feynman diagram calculations.

\section{The order $\veps$ dilatation operator}
\label{sec:dilatation_operator}

Let us consider a scalar field in $d$ dimensions with Lagrangian
\be 
\label{phi4_Lagrangian}
\Lm = \frac{1}{2} (\pd \phi)^2 + \frac{g}{4!}\phi^4\, .
\ee
It is well known that this theory has an interacting IR fixed point for dimensions $2 < d < 4$, and by considering the theory in $d=4-\veps$ dimensions it can be studied perturbatively. The $\beta$ function at order $\veps$ is
\be 
\beta(g) = - \veps g + \frac{3}{16\pi^2} g^2 + O(\veps^2)\, ,
\ee
which implies a weakly-coupled fixed point at
\be 
\label{g_star}
g_{*} = \frac{16\pi^2}{3}\veps + O(\veps^2)\, .
\ee
This is known as the Wilson-Fisher fixed point and its continuation to $d=3$ belongs to the same universality class as the three-dimensional critical Ising model. A systematic study of the spectrum of this system using Feynman diagrams was presented in \cite{Kehrein:1992fn,Kehrein:1994ff,Derkachov:1995wg,Kehrein:1995ia}. The approach in this paper is based on symmetry instead, where the conformal algebra will play a major role.

Local operators of the Wilson-Fisher fixed point are constructed by taking products of the fundamental scalar $\phi$ and its derivatives:
\be 
\Om(x) = \pd^{k_1} \phi(x) \pd^{k_2} \phi(x) \ldots \pd^{k_L} \phi(x)\, .
\ee
Borrowing terminology from the spin-chain literature we will sometimes call the individual fundamental fields ``letters''; the number of letters in a state is the ``length'' of the state. In this work we will be interested in the action of the \textit{dilatation operator} on the set of local operators. For $\veps=0$ the Wilson-Fisher fixed point is just free theory, this is the simplest example of a CFT where each field transforms under what we will call the zeroth-order conformal algebra. For $\veps \neq 0$ the fixed point becomes interacting with coupling \eqref{g_star}. This theory is still conformal and the conformal algebra of the free case will be modified by an interacting version. 
We will present the conformal algebra shortly, for the moment let us think of the conformal generators $J$ as a series expansion in $\veps$
\be 
J(\veps) = \sum_{k=0}^{\infty} J_k \veps^k\, .
\ee
For finite $\veps$ we have an interacting version of the conformal algebra where the commutation relations are unchanged, in particular, the scaling dimensions $\Delta$ of the generators stay the same:
\be 
[D(\veps),J(\veps)] = \Delta(J) J(\veps)\, ,
\ee
where $D$ is the dilatation operator. Our main goal is to obtain the lowest order correction to the dilatation operator, expanding the above identity we obtain,
\be
\label{interacting_scaling} 
[D(\veps),J(\veps)] = [D_0 + \veps D_1, J_0] + \veps [D_0,J_1] + O(\veps^2) = \Delta(J) J(\veps)\, ,
\ee
where the bare dimension of the generators remains unchanged \cite{Beisert:2003jj}
\be
\label{bare_dimension}
[D_0,J_k] = \Delta(J) J_k\, .
\ee
Plugging \eqref{bare_dimension} in \eqref{interacting_scaling} implies,
\be
\label{tree_level_invariance}
[J_0,D_1] = 0\, .
\ee
This identity and the ones above can be justified from perturbation theory. Following the approach of \cite{Rychkov:2015naa} we could organize them in an axiomatic way, however, we will not attempt to prove any rigorous theorems in this work. Our goal is simply to point out that symmetry plus intuition from perturbation theory is enough to fix the dilatation operator. Invariance under the zeroth-order conformal algebra \eqref{tree_level_invariance} significantly constrains the form of $D_1$. With the help of conformal representation theory one can fix its form up to an infinite set of coefficients $a(\ell)$, that are associated with the scaling dimension of conserved currents of spin $\ell$. 

\subsection{Representation theory of the conformal algebra}

The approach of the $\veps$ expansion is to consider the the Wilson-Fisher fixed point as a deformation of a CFT in $d=4$, therefore we will be concerned with the conformal algebra in four dimensions. The conformal algebra is generated by translations, special conformal transformations, Lorentz generators, and the dilatation operator:
\be 
\big\{ P_{\a \bd}\,, K^{\a \bd}\,, M^{\ph{\b}\a}_{\b}\,, \bar{M}^{\ph{\bd}\ad}_{\bd}\,, D \big\}\, .
\ee
Conformal multiplets are defined by a primary state with quantum numbers $(\Delta\,, j\,, \jb)$ that is killed by the $K$ generators:
\be
K^{\a \bd} |\Om_{\Delta, j, \jb} \rangle = 0\, ,
\ee
where $\Delta$ is the eigenvalue of $D$ and is known as the scaling or conformal dimension and $(j,\jb)$ label the Lorentz representation. Conformal descendants are then obtained by repeated application of the momentum operator
\be 
P^k |\Om_{\Delta, j, \jb} \rangle\, .
\ee
We should point out that by considering four-dimensional quantities we will not have access to the ``evanescent sector'' of the theory. These are operators that exist only in non-integer dimensions but that are nevertheless an integral part of the theory. This sector was studied recently in \cite{Hogervorst:2015akt}, where it was shown that the Wilson-Fisher fixed point is non-unitary in non-integer dimensions \footnote{For a bootstrap analysis in fractional dimensions see \cite{El-Showk:2013nia}}. In this paper our goal is to reproduce the earlier result by Kehrein et al. \cite{Kehrein:1992fn}, leaving the subtle but important issue of evanescent operators for future studies.

In a proper CFT any local operator should sit in an irreducible multiplet of the conformal algebra.
In general we expect generic long representations with Dynkin labels $(\Delta, j, \jb)$ that we denote as $\Am_{\Delta, j, \jb}$, the labels are not completely arbitrary and are subject to the following unitarity bounds:
\begin{align}
\begin{split}
\Delta \geqslant j + \jb + 2 \qquad & \text{for} \qquad  j \jb \neq 0\, ,
\\
\Delta \geqslant \jb + 1 \qquad & \text{for} \qquad  j = 0\, ,
\\
\Delta \geqslant j + 1 \qquad & \text{for} \qquad   \jb = 0\, .
\end{split}
\end{align}
There can also appear \textit{short multiplets} for some fixed values of the conformal dimension. These are usually associated to conservation equations. We denote the short multiplets of the conformal algebra using the following notation
\begin{align}
\begin{split}
\Cm_{j,\jb}: &  \qquad  \Delta = j + \jb + 2   ,
\\
\Bm^L_{j}: &  \qquad  \Delta = j + 1\,, \quad \jb=0\,,
\\
\Bm^R_{\jb}: &  \qquad  \Delta = \jb + 1\,, \quad j=0\,,
\\
\Bm: &  \qquad  \Delta = 1\,,  \quad j=\jb=0\,.
\end{split}
\end{align}
Some of these correspond to well known free-theory representations. For example, the multiplets of type $\Bm$, $\Bm^L_{\tfrac{1}{2}}$, $\Bm^R_{\tfrac{1}{2}}$ represent a free scalar, a free fermion, and a free anti-fermion respectively. The fact that the multiplet is short implies that one needs to mod out by a null state in the Verma module. Modding out by this null state translates into conservation equations in the field theory interpretation. Indeed, free theories satisfy
\begin{align}
\Bm: & \qquad \pd_{\a \ad} \pd^{\a \ad} \phi = 0\,,
\\
\Bm^L_{\tfrac{1}{2}}: & \qquad \pd^{\a \ad} \psi_{\a} = 0\,,
\\
\Bm^R_{\tfrac{1}{2}}: & \qquad \pd^{\a \ad} \bar{\psi}_{\ad} = 0\,.
\end{align}
Short multiplets that will play a key role in this work are the $\Cm_{j,j}$, these correspond to higher-spin conserved currents of spin $\ell=2j$:
\be 
\Cm_{j,j}: \qquad  \pd^{\a_1 \ad_1} J_{\a_1 \ldots \a_{2j},\ad_1 \ldots \ad_{2j}} = 0\,.
\ee
As already mentioned, in the Wilson-Fisher fixed point local operators can be built using the fundamental field $\phi$ and its derivatives $\pd^k \phi$, in representation theory language these fields belong to the same conformal multiplet $\Bm$: the field $\phi$ is the conformal primary and $\pd^k \phi$ its descendants.
A generic local operator $\Om$ can then be thought of as a product of $\Bm$ multiplets:
\be
\Om = \pd^{k_1} \phi \pd^{k_2} \phi \ldots \pd^{k_L} \phi \sim \Bm \times \Bm \times \ldots \times \Bm\, .
\ee

\subsection{The Hamiltonian density}

In the spin-chain literature it is customary to define a Hamiltonian by stripping off a power of the coupling from the dilatation operator 
\be 
D = \frac{g}{8\pi^2} H\, .
\ee
Here $D$ is the dilation operator and $g$ is the coupling constant. Using $g_{*}=\frac{16 \pi^2}{3} \veps$  at order $\veps$ the anomalous dimension and the energy eigenvalues $E$ of the Hamiltonian are related by
\be
\label{conversion} 
\gamma = \frac{2}{3} E\, .
\ee
In this paper we will use the terms energy and anomalous dimension interchangeably. 

At lowest order in the $\veps$-expansion the dilatation operator acts connecting pairs of letters in a state. This is natural from the Feynamn diagram origin of the corrections to anomalous dimensions. At order $\veps$ we have only one four-point vertex, which means that the first-order correction to the correlator
\be 
\langle \pd^{k_1}\phi(x_1) \phi \ldots \pd^{k_L}\phi(x_1) \pd^{k_1}\phi(x_2) \ldots \pd^{k_L}\phi(x_2) \rangle
\ee
will involve at most two interactions at a time. The action of the Hamiltonian can be written using a two-site Hamiltonian (or ``Hamiltonian density'')
\be 
\label{full_H}
H  = \sum_{i < j}^{L} H_{ij}\, , \qquad i,j= 1, \ldots L\, ,
\ee
where $L$ is the length of the operator. As before, this expression is justified from perturbation theory, but we can think of it as an axiom for the order $\veps$ correction to the dilatation operator. Our goal then is to calculate the density $H_{ij}$. Equation \eqref{tree_level_invariance} turns out to be very restrictive, indeed, invariance under the zeroth-order conformal algebra implies that $H_{ij}$ can be written as a sum of projectors into irreducible conformal multiplets. As already discussed, the fundamental field $\phi$ and its descendants belong to the $\Bm$ multiplet, the tensor product of two $\Bm$  can be written as an infinite sum of $\Cm$-type multiplets (see appendix \ref{app:characters})
\be 
\label{BtimesB}
\Bm \times \Bm = \sum_{\ell=0}^{\infty} \Cm_{\frac{\ell}{2},\frac{\ell}{2}}\, .
\ee
Because the Hamiltonian density is invariant under the zeroth-order conformal algebra, the above equation implies
\be 
\label{H_density}
H_{ij} = \sum_{\ell=0}^{\infty} a(\ell) P_{\ell}\, ,
\ee
where $P_{\ell}$ projects the product of two fundamental multiplets $\Bm \times \Bm$ to the multiplet $\Cm_{\frac{\ell}{2},\frac{\ell}{2}}$. As promised, conformal symmetry puts strong constraints on the form of the dilatation operator, which can be fixed up to an infinite set of coefficients $a(\ell)$. 

What remains now is to fix the $a(\ell)$ coefficients. One possibility, originally used in \cite{Beisert:2003jj}, is to restrict the Hamiltonian to a subsector with $SU(1,1)$ symmetry spanned by lightcone derivatives of the fundamental field $\pd_{+\dot{+}}^k\phi$ (see section \ref{sec:SU11_sector}). This a perturbative calculation, however, restricting the calculation to operators of the form $\pd_{+\dot{+}}^k\phi$ is a significant simplification when compared to the original calculation of \cite{Kehrein:1992fn}, where arbitrary operators were considered. In this sense, the expansion in \eqref{H_density} is already a success.
Thanks to recent results in the CFT literature we can actually do better. The multiplets $\Cm_{j,j}$ are associated to higher-spin currents, and this are known to have vanishing anomalous dimensions at order $\veps$, this is a well-known perturbative result, but it was recently re-derived using abstract CFT arguments \cite{Alday:2016njk,Alday:2016jfr,Roumpedakis:2016qcg}. Therefore, just from symmetry,
\be 
a(\ell) = 0\, , \qquad \ell > 0\, .
\ee
The only term that survives is $\ell=0$ which corresponds to the $\phi^2$ operator, but this anomalous dimension was fixed in the original paper by Rychkov and Tan \cite{Rychkov:2015naa}. The Hamiltonian then takes a very simple form (choosing $ij=12$)
\be 
\label{fixed_H}
H_{12} = \frac{1}{2}P_0\, .
\ee
As promised, we have fixed the order $\veps$ dilatation operator using only symmetry arguments. We will study the action of $P_0$ with more detail in the next section, for the moment let us point out that from the above expression and using \eqref{full_H} it is a trivial exercise to obtain
\be 
\gamma_{\phi^n} = \frac{1}{6}n(n-1)\, .
\ee

\section{The harmonic action}
\label{sec:harmonic_action}

In this section we present an efficient way to apply the Hamiltonian derived in the previous section, and calculate some sample anomalous dimensions as a check of our result.

\subsection{Oscillator representation}

To be able to act with the Hamiltonian \eqref{fixed_H} on an arbitrary state we use a modification of a simple formula found in \cite{Beisert:2003jj}.  It will be convenient to represent the states of the theory and the action of the conformal generators in terms of oscillators. Let us define two sets of bosonic oscillators ($\ab^{\a}$, $\ab^{\dagger}_{\a}$) and  ($\bb^{\ad}$,$\bb^{\dagger}_{\ad}$) with non-vanishing commutators given by
\begin{align}
[\ab^{\a},\ab^{\dagger}_{\b}] & = \delta^{\a}_{\b}\, ,
\\
[\bb^{\ad},\bb^{\dagger}_{\bd}] & = \delta^{\ad}_{\bd}\, .
\end{align}
The generators of the conformal algebra can be written as
\begin{align}
\nn
P_{\a \bd} & =  \ab^{\dag}_{\a} \bb^{\dag}_{\bd}\, , &
M^{\ph{\b}\a}_{\b}  & =  \ab^\dag_\b\ab^\a-\frac{1}{2}\delta^\a_\b \ab^\dag_\g\ab^\g\, ,
\\
K^{\a \bd} & = \ab^\a \bb^{\bd}\, ,
 &
\bar{M}^{\ph{\bd}\ad}_{\bd} & =  \bb^\dag_{\bd}\bb^{\ad}-\frac{1}{2}\delta^{\ad}_{\bd} \bb^\dag_{\gd}\bb^{\gd}\, , &
\\
\nn
D & =  1 + \frac{1}{2}\ab_{\g}^{\dag}\ab^\g + \frac{1}{2}\bb_{\gd}^{\dag}\bb^{\gd}\, , &
C & =   - \frac{1}{2}\ab_{\g}^{\dag}\ab^\g + \frac{1}{2}\bb_{\gd}^{\dag}\bb^{\gd}\, .
\end{align}
The generator $C$ is a central element that should kill valid physical states.
The fundamental field and its derivatives can be represented by creation operators acting on a vacuum state
\be 
\pd^k \phi = (\ab^{\dagger})^k(\bb^{\dagger})^k | 0 \rangle\, .
\ee
Introducing a collective oscillator $\textbf{A}_A=(\ab_{\ad},\bb_{\bd})$ an operator on $\Bm \times \Bm$ can be written in oscillator language as
\be 
\label{two_site_osc}
|s_1,...,s_n;A\rangle  = \textbf{A}^{\dagger}_{A_1,s_1} \ldots \textbf{A}^{\dagger}_{A_L,s_L}|00\rangle\, ,
\ee
where $s_{i}=1,2$ indicates the site where the oscillator sits (for simplicity we put the operators in position 1 and 2). In general, a valid state should satisfy the zero central charge constraint $C_1|A \rangle=0$ and $C_2|A \rangle=0$. 
To obtain the action of $H_{12}$, \eqref{fixed_H} tells us to project any two-site state into $\Cm_{0,0}$ and then multiply by $\tfrac{1}{2}$. In practice, this might be very cumbersome to implement, however, in \cite{Beisert:2003jj} a simple formula was presented, dubbed the ``harmonic action'', that makes the projection a straightforward calculation with oscillators. A simple modification of the harmonic action of $\Nm=4$ SYM will allow us to calculate the anomalous dimensions of the Wilson-Fisher fixed point.

The action of the Hamiltonian on a state \eqref{two_site_osc} should preserve the number of oscillators, but they can nevertheless change site. In general, we expect and expression of the form
\be 
\label{harmonic_action}
H_{12}|s_1,...,s_n;A\rangle = \sum_{s'_1,...,s'_n}c_{n,n_{12}+2,n_{21}}\delta_{C_1,0} \delta_{C_2,0}|s'_1,...,s'_n;A\rangle \, ,
\ee
The delta functions project onto states with zero central charge and $n_{ij}$ counts the number of oscillators moving from site $i$ to site $j$. 
The form of the coefficients $c$ has to be consistent with \eqref{fixed_H}. Their explicit form was worked out by Beisert for the $PSU(2,2|4)$ spin chain of $\Nm=4$ SYM. The appropriate formula for our case can be obtained from the full $PSU(2,2|4)$ formula by setting the fermions to zero. The explicit expression for the function $c_{n,n_{12},n_{21}}$ is
\be 
\label{cn_coefficients}
c_{n,n_{12},n_{21}}=\frac{1}{2}(-1)^{n_{12}n_{21}}\frac{\Gamma(\frac{1}{2}(n_{12}+n_{21}))\Gamma(1+\frac{1}{2}(n-n_{12}-n_{21}))}{\Gamma(1+\frac{1}{2}n)}\, ,
\ee
with $c_{n,0,0}=h(\frac{n}{2})$, where $h(j)$ is the $j$-th harmonic number. 

This concludes our implementation of the order $\veps$ dilatation operator. Although not obvious, the explicit formula for the action of $P_0$ is the one given in \eqref{harmonic_action} with the coefficients \eqref{cn_coefficients}. The proof follows the argument presented in \cite{Beisert:2003jj}. It can be checked that the harmonic action is invariant under the conformal algebra and that it has the correct normalization when acting on the $\Cm_{\frac{\ell}{2},\frac{\ell}{2}}$ multiplets. The harmonic action is very similar to the dilatation operator obtained in \cite{Kehrein:1992fn}, which was also described as a two-body interaction between states built using oscillators. The equality between our formula and theirs is not so transparent when looking at the explicit form of the dilatation operator, however, in the remainder of the paper we will diagonalize \eqref{harmonic_action} in some selected cases, and confirm that all our results are in perfect agreement with the literature.

\subsection{Some anomalous dimensions}

Armed with the complete order $\veps$ Hamiltonian we can now calculate anomalous dimensions for the primaries of the Wilson-Fisher fixed point. As a sample calculation, let us look at the low-lying primaries that appear in the OPE of $\phi^2$ with itself. The first four with at least four fields are (in schematic notation),
\be 
|\Pm_1 \rangle = \phi^4\, , \qquad |\Pm_2 \rangle = \Box^2 \phi^4\, , \qquad |\Pm_3 \rangle = \Box^3 \phi^4\, , 
\qquad |\Pm_4 \rangle = \Box^4 \phi^4\, .
\ee
with energies
\be 
E_1 = 3 \, , \qquad E_2 = \frac{5}{3}\, , \qquad E_3=  \frac{1}{2}\, , \qquad
E_4 =  \frac{7}{5}\, .
\ee
The explicit formulas for the primaries are,
\begin{align} 
| \Pm_1 \rangle = {} &  \phi^4
\\
\nn
\\
| \Pm_2 \rangle = {} &  \pd_\mu \pd_\nu \phi \pd_\mu \pd_\nu \phi \phi^2 -4 \pd_\mu \pd_\nu \phi \pd_\mu \phi \pd_\nu \phi \phi
+3 \pd_\mu \phi \pd_\mu \phi \pd_\nu \phi  \pd_\nu \phi
\\ 
\nn
\\
| \Pm_3 \rangle = {} &  \pd_\mu \pd_\nu \pd_\l \phi \pd_\mu \pd_\nu \pd_\l \phi  \phi^2 
-18\pd_\mu \pd_\nu \pd_\l \phi \pd_\mu \pd_\nu \phi \pd_\l  \phi \phi 
\\
\nn
&
+24 \pd_\mu \pd_\nu \pd_\l \phi \pd_\mu \phi \pd_\nu \phi \pd_\l \phi 
+63 \pd_\mu \pd_\nu \phi \pd_\mu \pd_\nu \phi \pd_\l \phi \pd_\l \phi 
\\
\nn
&
+24 \pd_\mu \pd_\nu \phi \pd_\mu \pd_\l \phi \pd_\nu \pd_\l  \phi \phi 
-108 \pd_\mu \pd_\nu \phi \pd_\mu \pd_\l \phi \pd_\nu \phi \pd_\l   \phi 
\\
\nn
\\ 
| \Pm_4 \rangle = {} &  \pd_\mu \pd_\nu \pd_\l \pd_\r \phi \pd_\mu \pd_\nu \pd_\l \pd_\r \phi  \phi^2 
-32 \pd_\mu \pd_\nu \pd_\l \pd_\r \phi  \pd_\mu \pd_\nu \pd_\l \phi \pd_\r \phi \phi 
\\
\nn
&  +36 \pd_\mu \pd_\nu \pd_\l \pd_\r \phi  \pd_\mu \pd_\nu \phi \pd_\l \pd_\r \phi \phi 
+64 \pd_\mu \pd_\nu \pd_\l \phi  \pd_\mu \pd_\nu \pd_\l \phi \pd_\r \phi \pd_\r \phi 
\\
\nn
& +144 \pd_\mu \pd_\nu \pd_\l \phi  \pd_\nu \pd_\l \pd_\r \phi \pd_\r \phi \pd_\mu \phi 
 -576 \pd_\mu \pd_\nu \pd_\l \phi  \pd_\mu \pd_\nu \phi \pd_\l \pd_\r \phi \pd_\r \phi 
\\
\nn
& +144 \pd_\mu \pd_\nu \pd_\l \phi  \pd_\nu \pd_\r \phi \pd_\r \pd_\l \phi \pd_\mu \phi 
+108 \pd_\mu \pd_\nu \phi  \pd_\nu \pd_\r \phi \pd_\r \pd_\l \phi \pd_\l \pd_\mu \phi 
\\
\nn
&
+117 \pd_\mu \pd_\nu \phi  \pd_\mu \pd_\nu \phi \pd_\l \pd_\r \phi \pd_\l \pd_\r \phi \, .
\end{align}
Applying the harmonic action is a straightforward algebraic exercise that can be efficiently implemented in a computer. In this paper we will not perform a systematic study of the order $\veps$ spectrum of the Wilson-Fisher fixed point, this was already done for operators up to length four in the original papers \cite{Kehrein:1992fn,Kehrein:1994ff,Kehrein:1995ia}, and for evanescent operators in \cite{Hogervorst:2015akt}. It might be worth revisiting the spectrum of this theory and compare with recent developments in the bootstrap. In particular, with the results of \cite{Simmons-Duffin:2016wlq} where the anomalous dimensions of more than 100 primaries were reported.

\subsection{Comments on multiplet recombination}

Let us now make some comments regarding multiplet recombination and higher-spin currents.\footnote{The analysis of this section was already discussed in \cite{Rychkov:2015naa}.} The higher-spin currents used to fix the Hamiltonian are the conformal primaries of the $\Cm_{j,j}$ multiplets of the conformal algebra. These multiplets satisfy the following decomposition rule:
\be
\label{recombination} 
\Cm_{j,j} = \Am_{2j+2,j,j}-  \Am_{2j+3,j-\frac{1}{2},j-\frac{1}{2}}\, .
\ee
For the stress-tensor multiplet $\Cm_{1,1}$ this implies
\be 
\Cm_{1,1} = \Am_{4,1,1} -  \Am_{5,\frac{1}{2},\frac{1}{2}}\, .
\ee
A recombination between $\Cm_{1,1}^4 $ and $\Am_{5,\frac{1}{2},\frac{1}{2}}$ in the interacting theory is not possible, because $\Am_{5,\frac{1}{2},\frac{1}{2}}$ is absent from the expansion of $\phi^4$ (this can be checked with the conformal characters presented in appendix \ref{app:characters}). Thus, the  stress-tensor multiplet remains short and therefore protected, as expected. 
For $\Cm_{2,2}$ the decomposition rule reads,
\be 
\Cm_{2,2} = \Am_{6,2,2} -  \Am_{7,\frac{3}{2},\frac{3}{2}}\, .
\ee
The multiplet $\Am_{7,\frac{3}{2},\frac{3}{2}}$ is present in the theory and could in principle recombine with the $\Cm$ multiplet to form a long multiplet. The fact that we found zero anomalous dimension for all the higher-spin currents at one-loop, puts tight restrictions in the value of the anomalous dimension of the $\Am_{7,\frac{3}{2},\frac{3}{2}}$ multiplet.
The conformal dimensions of $\Cm_{2,2}$ and $\Am_{7,\frac{3}{2},\frac{3}{2}}$  are
\be 
\label{dims}
d-2 + 4\, , \qquad 2d - 4 + 3 + \gamma \veps\, ,
\ee
where the anomalous dimension $\gamma$ can be calculated using our Hamiltonian. In order for these multiplets to recombine, the value of $\gamma$ should be such that the difference between the conformal dimensions is exactly one. If this did not happen, it would imply that recombination is not possible and the $\Cm$ multiplet will remain short at all orders. Motivated by this we proceed to calculate the order $\veps$ anomalous dimension of the $\Am_{7,\frac{3}{2},\frac{3}{2}}$ multiplet. The primary in schematic form is,
\be 
| \Pm \rangle = \pd^3 \phi \phi^3 + 9 \pd^2 \phi \pd \phi \phi^2 + 12(\pd \phi)^3 \phi\, ,
\ee
and the action of the Hamiltonian gives,
\be 
H | \Pm \rangle = \frac{3}{2} | \Pm \rangle\, .
\ee
Using the conversion formula \eqref{conversion} and replacing in (\ref{dims}) we obtain,
\be 
6 - \veps\, , \qquad 7 -\veps \, .
\ee
We see that the value of $\gamma$ is such that the difference between the two scaling dimensions is exactly one.
The same question can be asked about the $\Cm_{3,3}$ and $\Am_{9,\frac{5}{2},\frac{5}{2}}$ multiplets, applying the Hamiltonian on the corresponding $\pd^5 \phi^4$ primary the eigenvalue is again $\frac{3}{2}$. For the  $\Cm_{4,4}$ and $\Am_{11,\frac{7}{2},\frac{7}{2}}$ case there are two $\pd^7 \phi^4$ primaries with eigenvalues ($\frac{1}{2},\frac{3}{2}$). The conclusion is again the same, the multiplet with energy $\frac{3}{2}$ can recombine with the $\Cm$ multiplet to form a long multiplet. Our results seem to indicate that a similar phenomenon will occur for higher spins: there will always be at least one multiplet with energy $\frac{3}{2}$.

\section{The $SU(1,1)$ subsector}
\label{sec:SU11_sector}

In this section we present an alternative way to calculate the dilatation operator of the Wilson-Fisher fixed point. Like in the previous sections, most of the techniques used here were borrowed from the spin-chain literature, in particular the work of Beisert on the $PSU(2,2|4)$ spin-chain of $\Nm=4$ SYM \cite{Beisert:2003jj,Beisert:2004ry}. In the previous section we managed to fix all the $a(\ell)$ coefficients invoking CFT arguments, but this might not always work.  Beisert's approach is quite universal and should be applicable to more general theories that admit an $\veps$ expansion. If in a more complicated model some of the $a(\ell)$ coefficients remain unfixed, a perturbative calculation might be necessary. For the Wilson-Fisher fixed point this sector was studied to second order in $\veps$ in \cite{Kehrein:1995ia}.

\subsection{Lifting the Hamiltonian}

The starting point is again the two-site Hamiltonian
\be 
\label{two_site_H}
H_{12} = \sum_{\ell=0}^{\infty}a(\ell) P_{\ell}\, ,
\ee
where $P_{\ell}$ projects a two-site state into its irreducible  $\Cm_{(\frac{\ell}{2},\frac{\ell}{2})}$ component. In the remainder of this section we will calculate the $a(\ell)$ coefficients by restricting the Hamiltonian to a special subsector with $SU(1,1)$ symmetry. Let us consider the subsector spanned by lightcone derivatives of the scalar field
\be 
\label{SU11_letter}
\pd_{+ \dot{+}}^k \phi\, .
\ee
Operators are then built out of an arbitrary number of fields with an arbitrary number of derivatives acting on them. 

The fundamental letter \eqref{SU11_letter} of the sector corresponds to an infinite $SU(1,1)$ representation of ``spin'' $n=-\tfrac{1}{2}$. Taking the product of two spin $n=-\tfrac{1}{2}$ representations one can easily prove (see appendix \ref{app:characters})
\be 
\label{SU11_tensor}
V_{-\tfrac{1}{2}} \times V_{-\tfrac{1}{2}} = \sum_{n=0}^{\infty} V_{-1-n}\, .
\ee
The Hamiltonian in this subsector is an $SU(1,1)$ invariant and following the same arguments of section \ref{sec:dilatation_operator} one can write it as an infinite sum of $SU(1,1)$ projectors
\be 
H^{'}_{12} = \sum_{n=0}^{\infty}b(n)V_{-1-n}\, .
\ee
It turns that there is a one-to-one correspondence between the conformal $\Cm_{(\frac{\ell}{2},\frac{\ell}{2})}$ multiplets and the $SU(1,1)$ $V_{-1-n}$ multiplets. This implies that we can obtain the coefficients $a(\ell) \equiv b(n)$ of \eqref{two_site_H} just by studying this restricted subsector. This ``lifting'' of the Hamiltonian was the trick used in \cite{Beisert:2003jj,Beisert:2004ry} to obtain the complete one-loop dilatation operator of $\Nm=4$ SYM. Similar methods were used to obtain the one-loop dilatation operator of ABJM \cite{Zwiebel:2009vb}, $\Nm=2$ and $\Nm=1$ SCQCD \cite{Liendo:2011xb,Liendo:2011wc}, and $\beta$-deformed $\Nm=4$ SYM \cite{Fokken:2013mza}.

At order $\veps$ we only need to concentrate on states with two letters
\be 
\frac{\pd_{+\dot{+}}^k \phi}{k!}\frac{\pd_{+\dot{+}}^{n-k}\phi}{(n-k)!}\, .
\ee
The action of the Hamiltonian on this state can be calculated using standard perturbation theory and it was originally obtained in \cite{Kehrein:1994ff,Derkachov:1995wg}:
\be
\label{Htw2} 
H^{'}_{12}\frac{\pd_{+\dot{+}}^k \phi}{k!}\frac{\pd_{+\dot{+}}^{n-k}\phi}{(n-k)!} = \sum_{k'=0}^n \frac{1}{2(n+1)}\frac{\pd_{+\dot{+}}^{k'} \phi}{k'!}\frac{\pd_{+\dot{+}}^{n-k'}\phi}{(n-k')!}\, .
\ee
From (\ref{Htw2}) we can now obtain the anomalous dimensions of the higher-spin currents,
\be 
\label{currents}
J_{n} = \sum_{i=0}^n(-1)^i\binom{n}{i}\frac{\pd_{+\dot{+}}^{i} \phi}{i!}\frac{\pd_{+\dot{+}}^{n-i}\phi}{(n-i)!}\,.
\ee
Using the oscillator representation it is not hard to check that these currents are the $SU(1,1)$ primaries of the $V_{-1-n}$ multiplets. It can also be checked that they are the conformal primaries of the $\Cm_{\frac{n}{2},\frac{n}{2}}$ multiplets.
Acting with the explicit $SU(1,1)$ Hamiltonian \eqref{Htw2} we obtain\footnote{Actually, only currents with even spin appear in this expansion, currents with odd spin are automatically zero due to Bose symmetry. However, one can prove that the action of the Hamiltonian on odd spin currents is zero without requiring Bose symmetry, for example, $H_{12}(\pd_{+\dot{+}}\phi \phi - \phi \pd_{+\dot{+}}\phi) = 0$. This is relevant if the fields $\phi$ have extra indices, like in the $O(N)$ model.},
\be
H^{'}_{12} J_n = \frac{1}{2}\delta_{n,0}J_n\, .
\ee
The only term that survives is the $J_0$ primary which corresponds to the $\phi^2$ operator. If we multiply the energy eigenvalue by the $2/3$ conversion factor given in \eqref{conversion} we obtain
\be 
\gamma_{\phi^2} = \frac{1}{6}\, ,
\ee
which is indeed the right value for this anomalous dimension. The higher-spin currents on the other hand are all conserved at order $\veps$, as expected. It is well known that these currents acquire anomalous dimensions at order $\veps^2$. The sole exception being the $J_1$ current, which corresponds to the spin two stress tensor that remains protected at all orders. 
The fact that all the currents have zero order $\veps$ anomalous dimension implies that the only non-zero coefficient in \eqref{two_site_H} is $a(0)=\frac{1}{2}$. Therefore, the Hamiltonian of the full theory, not necessarily restricted to the $SU(1,1)$ sector, is
\be 
H_{12} = \frac{1}{2} P_0 \, .
\ee
The same result obtained in section \ref{sec:dilatation_operator}

\subsection{Anomalous dimensions for length $L=3$}

As a final check of our approach let us now diagonalize the Hamiltonian in the $SU(1,1)$ sector for operators with $L=3$ using the harmonic action of section \ref{sec:harmonic_action}. The procedure is completely algorithmic, we organize the operators in order of increasing spin $\ell$, and for each spin we list a basis of operators given by all inequivalent combinations of derivatives acting on the fields $\phi$:
\begin{align}
\vec{B}_{\ell=0} : & \qquad \big\{|\phi \phi \phi \rangle \big\}\, ,
\\
\vec{B}_{\ell=1} : & \qquad \big\{|\phi\phi\pd_{+\dot{+}} \phi\rangle \big\}\, ,
\\
\vec{B}_{\ell=2} : & \qquad \big\{|\phi \pd_{+\dot{+}}\phi \pd_{+\dot{+}}\phi\rangle, 
|\phi \phi\pd_{+\dot{+}}^2 \phi\rangle \big\}\, ,
\\
\vdots
\end{align}
Once the basis states are listed we need to obtain conformal primaries. To achieve this we consider the most general linear combination of basis vectors with arbitrary coefficients $\a_i$ and impose
\be 
\label{K_action}
(K^{+\dot{+}}_1 + K^{+\dot{+}}_2 + K^{+\dot{+}}_3) \sum_{i=0}^{\text{dim} \vec{B}_{\ell}} \a_i B_{\ell}^i = 0\, .
\ee
For low spins there is only one solution to this condition. Let us call this family of primaries $|\Pm_{1,\ell}, L=3\rangle$. By applying our Hamiltonian we can obtain their energies
\begin{align}
H |\Pm_{1,\ell}, L=3\rangle & = \left(\frac{1}{2}-\frac{1}{1+\ell} \right) |\Pm_{1,\ell}, L=3\rangle\, ,
\qquad \ell\, \, \text{odd}
\\
H |\Pm_{1,\ell}, L=3\rangle & = \left(\frac{1}{2}+\frac{1}{1+\ell} \right) |\Pm_{1,\ell}, L=3\rangle\, .
\qquad \ell\, \, \text{even}
\end{align}
Starting with $\ell=6$ for $\ell$ even and $\ell=9$ for $\ell$ odd a new family of primaries appear. Let us denote this second family by $|\Pm_{2,\ell}, L=3\rangle$. Acting with the Hamiltonian on these states we obtain
\be 
H |\Pm_{2,\ell}, L=3\rangle = 0\, ,
\ee
and they are therefore protected at order $\veps$.

More and more primaries will appear with increasing spin making the analysis more demanding, so we will not consider them in this work. The actual number of primaries can be obtained using the characters presented in appendix \ref{app:characters}. To take into account Bose symmetry we need to use the plethystic exponential
\be 
\label{PE}
\textrm{PE}(z C_{-\tfrac{1}{2}}(x)) = \exp{\left(\sum_{n=1}^\infty\frac{z^n}{n} C_{-\tfrac{1}{2}}(x^n)\right)}\, ,
\ee
where $z$ labels the number of fundamental letters of spin $j=-\frac{1}{2}$. The coefficient of the $z^3$ term in (\ref{PE}) is
\be
\chi^{(3)}  =  \frac{1}{6}\bigg(C_{-\tfrac{1}{2}}(x) + 3 C_{-\tfrac{1}{2}}(x)C_{-\tfrac{1}{2}}(x^2)
+2 C_{-\tfrac{1}{2}}(x^3)\bigg)\, .
\ee
Here we have defined the character  $\chi^{(3)}$ associated with three Bose symmetrized spin $n=-\tfrac{1}{2}$ representation.
We can decomposed $\chi^{(3)}$ as an infinite sum of $V_{n}$ multiplets and from there read the multiplicity of the primaries at each spin. It is not hard to check that the generating functional for this counting is
\be 
\sum_{\ell=0}^{\ell} c_{\ell} x^{3\ell} = \frac{1}{(1-x^6)(1-x^9)}\, .
\ee
The spin dependence of these anomalous dimensions was already known in \cite{Kehrein:1994ff,Derkachov:1995wg}. The same result was obtained in \cite{Roumpedakis:2016qcg} where operators where the twist equals the length where studied using multiplet recombination arguments. These type of operators are precisely the ones captured by the $SU(1,1)$ subsector where the spin and free conformal dimension of the operators are constrained by
\be 
\Delta_{\text{free}}-\ell = L\, .
\ee
It is also not surprising that our counting of primaries using $SU(1,1)$ characters coincides with the counting of \cite{Roumpedakis:2016qcg}.

\subsection{Dilatation operator of the $O(N)$ model}

As a final application let us generalize our results for the more general $O(N)$ model. In this case we have a scalar with an $O(N)$ vector index $\phi^i$ described by the Lagrangian,
\be 
\label{ON_lagrangian}
\Lm = \frac{1}{2} (\pd \phi^i)(\pd \phi^i) + \frac{g}{4}(\phi^i\phi^i)^2\, .
\ee
and the IR perturbative fixed point is given by
\be 
g_{*} = \frac{8\pi^2}{N+8}\veps + O(\veps^2)\, .
\ee
The $O(N)$ flavor symmetry commutes with the conformal algebra and therefore all the considerations of the previous sections apply. The only difference is that now we have extra flavor indices along for the ride. In order to obtain the Hamiltonian it is enough to fix it in the $SU(1,1)$ sector. Its action now reads, 
\be
\label{Htw2ON} 
H_{12}\frac{\pd^k \phi_i}{k!}\frac{\pd^{n-k}\phi_j}{(n-k)!}  =  \frac{\delta_{ij}}{3}\sum_{k'=0}^n \frac{1}{2(n+1)}\frac{\pd^{k'} \phi_l}{k'!}\frac{\pd^{n-k'}\phi_l}{(n-k')!}
+ \frac{2}{3}\sum_{k'=0}^n \frac{1}{2(n+1)}\frac{\pd^{k'} \phi_i}{k'!}\frac{\pd^{n-k'}\phi_j}{(n-k')!}\, .
\ee
It is evident that the two terms in this equation are almost identical to the ones for the scalar theory. The only differences being an overall coefficient and the presence of flavor structure. Thus, the Hamiltonian for the $O(N)$ model is just
\be 
H_{12} = \frac{2}{3}P_0 \mathbb{I} + \frac{1}{3}P_0 \mathbb{K}\, ,
\ee
where $\mathbb{I}$ and $\mathbb{K}$ are the identity and trace operators in flavor space:
\be 
\mathbb{I} \phi^i \phi^j = \phi^i \phi^j\, , \qquad \mathbb{K} \phi^i \phi^j = \delta^{ij} \phi^k \phi^k\, .
\ee
The harmonic action is the same as before and the calculation of anomalous dimensions proceeds in the same way.

Let us check that our previous comments about multiplet recombination also applies for this theory. The unique $\pd^3 \phi^4$ primary in the $O(N)$ case is (schematically),
\be
| \Pm \rangle  =  \pd^3 \phi_i \phi_i \phi_j \phi_j - 3 \pd^2 \phi_i \pd \phi_i \phi_j \phi_j
 - 6 \pd^2 \phi_i  \phi_i \pd \phi_j \phi_j + 12\pd \phi_i \pd \phi_i \pd \phi_j \phi_j \, ,
\ee
which corresponds to a long multiplet $\Am_{7,\tfrac{3}{2},\tfrac{3}{2}}$. The action of the Hamiltonian gives
\be 
H | \Pm \rangle = \frac{N+8}{6} | \Pm \rangle\, ,
\ee
which is again the correct value such that recombination with a $\Cm_{2,2}$ multiplet is possible.

\section{Conclusions and future directions}
\label{sec:conclusions}

In this work we have obtained the dilatation operator of the Wilson-Fisher fixed point using symmetry principles. There were two key steps in the derivation. First, conformal invariance of the dilatation operator fixes its form up to an infinite set of coefficients. Second, the unfixed coefficients have been recently obtained using CFFT arguments. The combination of both approaches gave the desired result reproducing the Feynman diagram calculation of \cite{Kehrein:1992fn}.

Similar techniques should be applicable to more general conformal fixed points that admit an $\veps$ expansion. Examples that have been studied using multiplet recombination are the Gross-Neveu model in $d=2+\veps$, $\phi^3$ in $d=6-\veps$, and $\phi^6$ in $d=3-\veps$ \cite{Ghosh:2015opa,Raju:2015fza,Basu:2015gpa,Nii:2016lpa} (see also \cite{Hasegawa:2016piv}). Models that have been studied recently using standard $\veps$-expansion perturbation theory are QED$_3$ \cite{DiPietro:2015taa}, and conjectural fixed points in dimensions higher than four \cite{Fei:2014yja,Fei:2015kta,Stergiou:2015roa,Pang:2016xno,Mati:2014xma,Mati:2016wjn}. In these models the fundamental fields include fermions and gauge fields, fields with Lorentz indices have already appeared in the calculation of the dilatation operator in supersymmetric theories \cite{Zwiebel:2009vb,Liendo:2011xb,Liendo:2011wc,Fokken:2013mza}, and most likely many of those results can be recycled for the non-supersymmetric cases. Finally, it might also be interesting to use this approach to study CFTs in the presence of defects \cite{Gaiotto:2013nva,Gliozzi:2015qsa,Chester:2015wao}.

Another venue is to try to fix higher-order corrections to the dilatation operator. A two-loop analysis in the $SU(1,1)$ subsector was already presented in \cite{Kehrein:1995ia}. It is possible that this result contains ll the unknown coefficients not fixed by symmetry. Moreover, order $\veps^2$ results have also been obtained using symmetry arguments, so there is a fighting chance of fixing the second order correction without relying on perturbation theory. In $\Nm=4$ SYM where the operator content is far more complicated, the two-loop dilatation operator has been an open problem for many years, the state of the art being \cite{Zwiebel:2011bx}.

Perhaps the most interesting future direction is to generalize the analysis to the evanescent sector studied in \cite{Hogervorst:2015akt}. However, it seems that this requires the development of new technology. All the conformal algebra identities used in this work were four-dimensional. Generalizing these expressions to real $d$ and working out the corresponding character technology seems to be a necessary first step.


\section*{Acknowledgments}
I have benefited from discussions with M.~Lemos, E.~Pomoni, L.~Rastelli, S.~Rychkov, V.~Schomerus, and B.~van Rees.

\appendix
\addtocontents{toc}{\protect\setcounter{tocdepth}{1}}
\section{Characters}
\label{app:characters}

In this appendix we collect character identities useful for proving some of the equations in the main text. See for example \cite{Dolan:2005wy} where the origin of the formulas presented here is explained.

\subsection{$SU(1,1)$ characters}

The $SU(1,1)$ character for a representation of spin $j$ is given by
\be 
\label{su11_character}
C_{j}(x) = \frac{x^{2j+1}}{x-x^{-1}}\, .
\ee
The fundamental letters $\pd_{+\dot{+}}^k \phi$ transform in the spin $j=-\frac{1}{2}$ representation, using \eqref{su11_character} it is straightforward to check identity
\be 
\left(C_{-\tfrac{1}{2}}(x)\right)^2 = \sum_{j=0}^{\infty} C_{-1-j}(x)\, ,
\ee
confirming the tensor product expansion \eqref{SU11_tensor}.

\subsection{Conformal characters}

The conformal characters for the $\Bm$, $\Cm_{j,\jb}$, and $\Am_{\Delta,j,\jb}$ multiplets are
\begin{align}
\chi_{\Bm} & = s^2(1-s^4)P(s,x,y)\, ,
\\
\\
\chi_{\Cm} & =s^{2 j + 2 \jb + 4} \left(\chi_j(x) \chi_{\jb}(y) - 
   s^2 \chi_{j - \tfrac{1}{2}}(x) \chi_{\jb - \tfrac{1}{2}}(y)\right) P(s, x, y)\, ,
\\
\chi_{\Am} & = s^{2 \Delta}\chi_j(x)\chi_{\jb}(y)P(s,x,y)\, .
\end{align}
Here $\chi_j(x)$ is an $SU(2)$ character
\be 
\chi_j(x) = \frac{x^{2j+1}-x^{-2j-1}}{x-x^{-1}}\, ,
\ee
and $P(s,x,y)$ captures the conformal descendants obtained by repeated action of the momentum operator
\be 
P(s,x,y) = \frac{1}{(1-s^2 x y)(1-\frac{x}{y}s^2 )(1-\frac{y}{x}s^2 )(1- \frac{s^2}{x y})}\, .
\ee
The conformal dimension $\Delta$ is labeled by $s$ and the left and right Lorentz quantum numbers $(j,\jb)$
are labeled by $(x,y)$. Similarly to the $SU(1,1)$ case, the tensor product relation \eqref{BtimesB} is a consequence of the character identity
\be 
\left( s^2(1-s^4)P(s,x,y) \right)^2 = \sum_{\ell=0}^{\infty} s^{2 \ell + 4} \left(\chi_{\tfrac{\ell}{2}}(x) \chi_{\tfrac{\ell}{2}}(y) - 
   s^2 \chi_{\tfrac{\ell-1}{2}}(x) \chi_{\tfrac{\ell-1}{2}}(y)\right) P(s, x, y)\, .
\ee
The decomposition rule \eqref{recombination} can also be easily checked with characters.

\newpage

\bibliographystyle{abe}
\bibliography{bibliography}

\end{document}